\documentclass[twocolumn]{svjour3}          
\smartqed  
\usepackage{graphicx}
\usepackage{dcolumn}
\usepackage{bm}
\usepackage{hyperref}

\usepackage{cite}

\begin{document}

\title{Liquid soap film generates electricity}

\subtitle{A suspended liquid film rotating in an external electric field as an electric generator}


\author{A. Amjadi        \and
        M. S. Feiz     	 \and
        R. M. Namin.
}


\institute{A. Amjadi \at
              Physics Department, Sharif University of Technology, Tehran, Iran. \\
              \email{amjadi@sharif.edu}           
           \and
           M. S. Feiz \at
              Physics Department, Sharif University of Technology, Tehran, Iran. 
           \and
           R. M. Namin \at
              Department of Mechanical Engineering, Sharif University of Technology, Tehran, Iran.
}

\date{Accepted for publication in \textbf{Microfluidics and Nanofluidics} \\
Received: 08 October 2013 / Accepted: 23 April 2014 \\
\textit{Brief Communication}
}


\maketitle

\begin{abstract}
We have observed that a rotating liquid soap film generates electricity when placed between two non-contact electrodes with a sufficiently large potential difference.
In our experiments suspended liquid film (water + soap film) is formed on the surface of a circular frame, which is forced to rotate in the $x-y$ horizontal plane by a motor. This system is located at the center of two capacitor-like vertical plates to apply an external electric voltage difference in the $x-$direction. The produced electric current is collected from the liquid film using two conducting electrodes that are separated  in the $y-$direction.
We previously reported that a liquid film in an external electric field rotates
when an electric current passes through it, naming it the liquid film motor (LFM).
In this paper we report a novel technique, in which a similar device can be used as an electric generator,
converting the rotating mechanical energy to electrical energy.
The liquid film electric generator (LFEG) is in stark contrast to the LFM, both of which could be
designed similarly in very small scales like micro scales with different applications. Although the device is comparable to commercial electric motors or electric generators, there is a significant difference in their working principles. Usually in an electric motor or generator the magnetic field causes the driving force, while in a LFM or LFEG the Coulomb force is the driving force. This fact is also interesting from the Bio-science point of view and brings a similarity to bio motors.
Here we have investigated the electrical characteristics of such a generator for the first time experimentally and modelled the phenomenon with electroconvection governing equations. A numerical simulation is performed using the local approximation for the charge-potential relation and results are in qualitative agreement with experiments.
\end{abstract}

\section{Introduction}

The interaction of electric fields with liquids, known as Electrohydrodynamics, has been a subject of research for more than fifty years. Many different phenomenon are studied extensively in this field, including
Taylor cones and electrospray \cite{taylor1964disintegration,chen2005spraying,marginean2006much,fernandez2007fluid,collins2007electrohydrodynamic},
electrohydrodynamic whipping jets \cite{taylor1969electrically,hohman2001electrospinning,riboux2011whipping},
electrowetting and anti-coalescent drops \cite{walker2009electrowetting,bird2009critical},
liquid bridges \cite{burcham2000electrohydrodynamic,burcham2002electrohydrodynamic,namin2013experimental},
etc. These phenomenon are explained and modelled with the Leaky-Dielectric model developed by Taylor \&  Melcher in the 1960s \cite{melcher1969electrohydrodynamics,saville1997electrohydrodynamics}.

Electric motors and generators, as devices which convert electrical energy to mechanical energy and vice versa, have been studied and developed for more than a century
\cite{hertz1881vertheilung,quincke1896ueber}.
Different mechanisms have been developed for this purpose, working with AC and DC voltages.
The application of electrohydrodynamic mechanisms for such systems becomes interesting in many fields including micro total analysis systems ($\mu$TAS) and Lab-on-a-Chip (LOC) for use in biological and chemical assays
\cite{zeng2004principles,janasek2006scaling,chang2007electrokinetic,chang2010electrokinetically,ramos2011electrokinetics,lee2011microfluidic,zhao2012advances}.
One reason for this is that they enable a device with no external mechanical moving parts and can be made portable using batteries. In 1955 Sumoto \cite{sumoto1955interesting} reported a device that rotates a shaft in a liquid by the application of a DC high voltage. More recently, Sugiyama \textit{et al.} \cite{sugiyama2008development} constructed an EHD motor that could rotate a rotor immersed in a dielectric fluid by a DC field.
The presence of a solid rotor might be a limitation for such motors, and if one could rotate a fluid with no external mechanical moving parts, this might come useful in the micro scales e.g. for micro-mixing or washing. In 2013, Salipante \& Vlahovska \cite{salipante2013ehd} reported and theoretically explained the deformation and rotation of a droplet in a uniform electric field.
In a leaky dielectric, because of the conductivity of the fluid, electric charge density is zero except for a very thin layer near the interfaces of the liquid. Thus the electric force is exerted to a region very close to the interfaces. 
If only solid-liquid interfaces exist, because of the no-slip condition, a very weak motion is induced. So liquid-gas interfaces will be more effective, and this is why a droplet can show special deformation and rotational behaviour. \cite{salipante2013ehd}
An extreme condition in this case is a suspended thin liquid film, where very large interfaces encounter with air. Electrohydrodynamics of thin films have been extensively studied for the case of Electroconvection experimentally \cite{tsai2004aspect}, theoretically \cite{deyirmenjian1997weakly} and numerically \cite{tsai2007direct}. When suspended thin films are placed in electric fields, electric charges are accumulated on the free surfaces of the liquid gas interface. These charges are subject to an electrical force in presence of an electric field. This force is the main mechanism behind electroconvection, which is an instability caused when current passes through a film.

In this context, we have previously introduced a device called the liquid film motor (LFM) \cite{amjadi2009, shirsavar2012rotational, amjadi2013electro}. The device was consisted of a horizontal two dimensional frame on which a suspended liquid film is formed, connected to a pair of conducting electrodes which conduct an electric current through it. Two parallel vertical metal plates are placed on two sides of the film and are connected to a DC high voltage. When voltages are connected to the electrodes and the plates, the film starts to rotate.
To explain the physical mechanism for this rotation, Shiryavea \textit{et al.}\cite{shiryaeva2009theory} in 2009 performed calculations using classical Electrohydrodynamics, although the surface charge mechanism explained above and known from the literature on electroconvection was missed out.

\begin{figure}[b]
 \includegraphics[width=\linewidth]{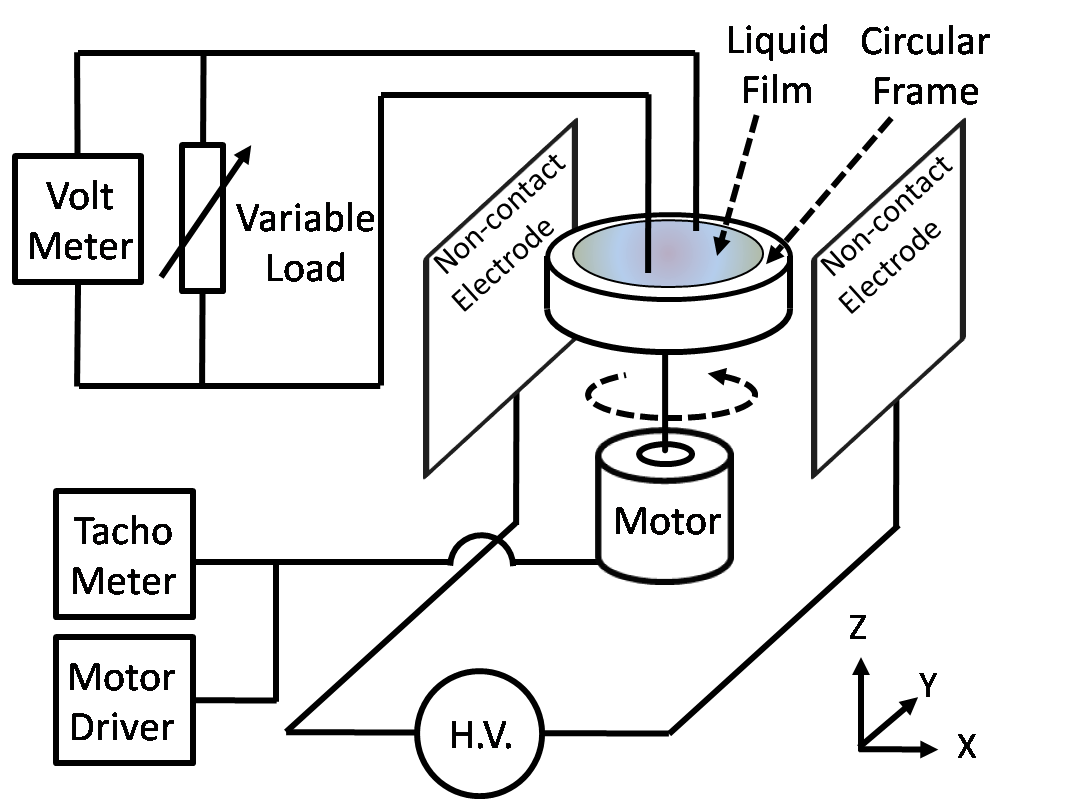}
 \caption{Schematic of the experimental setup.}
 \label{fig:pic}
\end{figure}

In this letter we construct a different experiment, treating the previous liquid film differently. The device behaves as a generator, i.e. it converts the rotating mechanical energy to electrical energy.
The Liquid Film Electric Generator (LFEG) behaves the inverse of LFM, i.e., If we apply an electric current on a suspended liquid film placed between two non-conducting electrodes it rotates,
and on the other hand, if we force the liquid film to rotate, it generates electricity.

\section{Experiments}
The setup consists of a circular frame attached and centred to an electric motor to be rotated (Fig. \ref{fig:pic}). The frame diameter is $20 mm$. A liquid soap film is formed on the frame and two wires of diameter 10$\mu m$ are dipped $14 mm$ apart into the film as electrodes to conduct the current through. The film was placed between two vertical metal plates (the non-conduct electrodes), $l_{ext} = 6 cm$ apart, which were connected to a high voltage DC power supply of a few  kilo Volts. One plate was connected to a voltage of $V_{ext}/2$ and the other connected to $-V_{ext}/2$.
When the voltage was applied to the non-contact electrodes and the film was forced to rotate, a potential difference and electric current could be measured through the conducting electrodes.

The thickness of the film varies with its angular velocity due to centrifugal effects. To control this, in every experiment the film was first rotated at the maximum angular velocity ($4000 RPM$), then it was slowed to perform the measurements. As a result, the variations in thickness were reduced. The average thickness of the film, estimated from the reflection color, was approximately between 200 to 500 nanometers in all of the experiments.

\begin{figure}
\centering
\includegraphics[width = 0.35\linewidth]{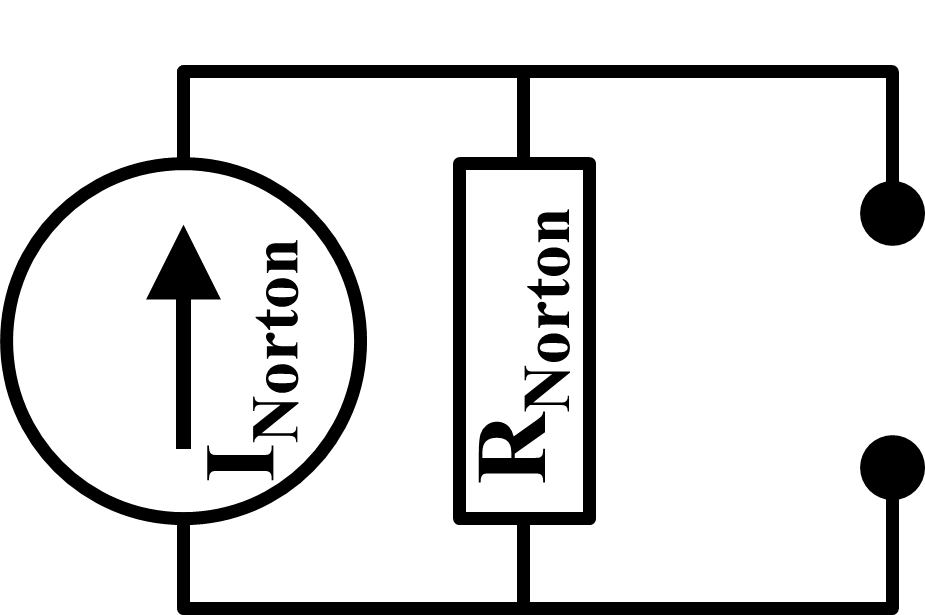} \ \ \ \ 
\includegraphics[width = 0.35\linewidth]{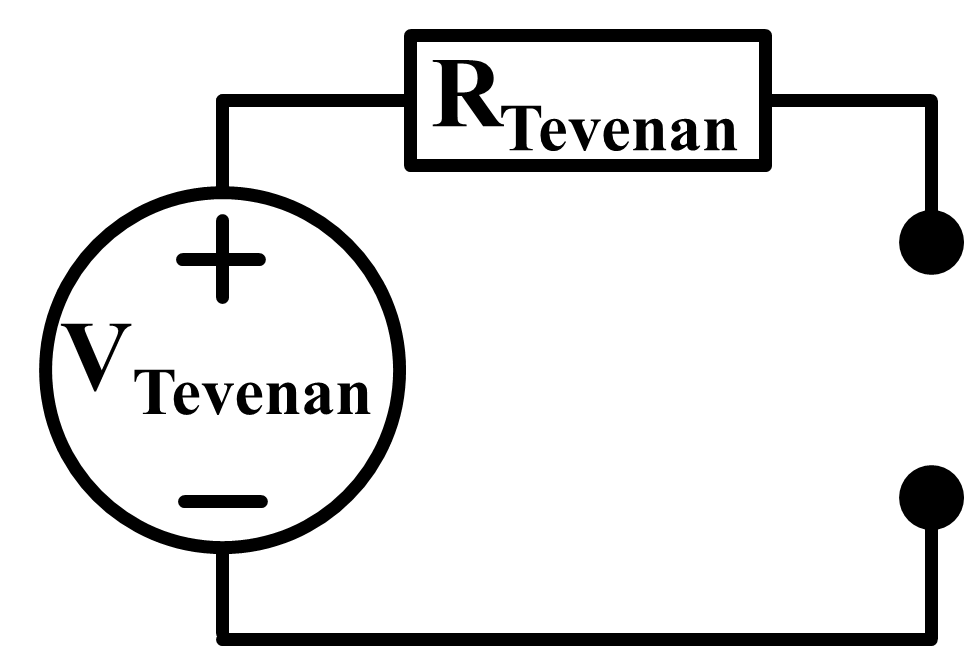}
\caption{Norton (left) and Tevenan (right) equivalents of the generator.}
\label{fig:nort}
\end{figure}

\begin{figure}[h]
 \includegraphics[width=80mm]{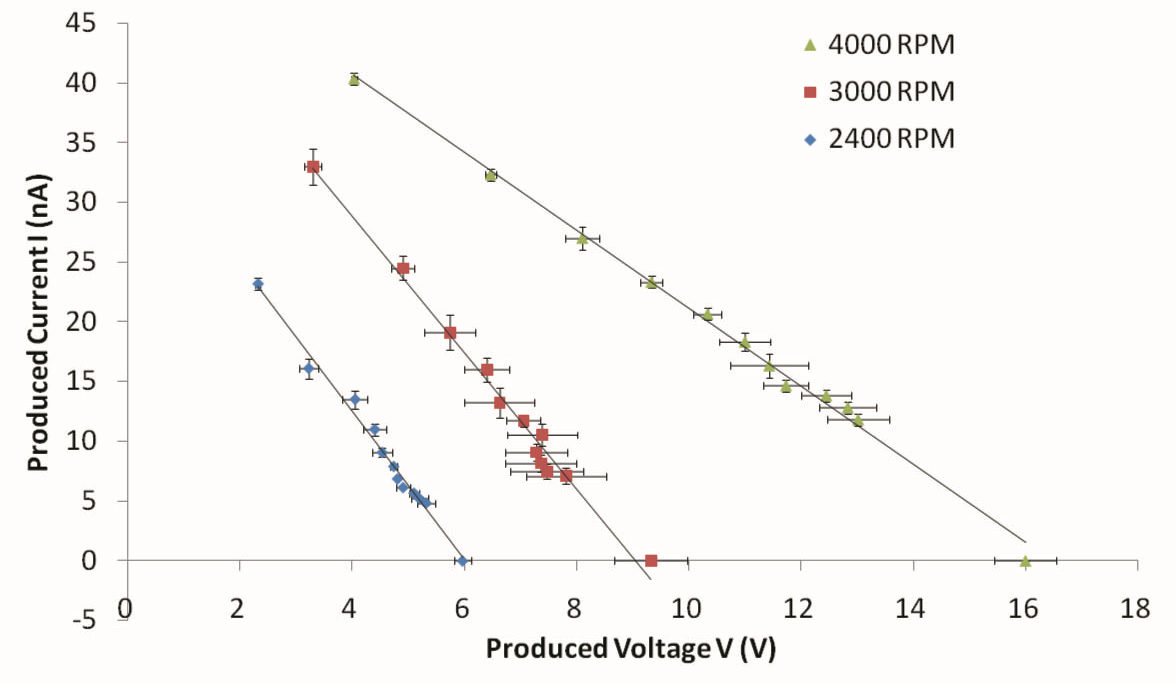}
 \caption{$I-V$ Characteristics of the liquid film electric generator in different angular velocities. $V_{ext} = 6kV$.
 	Solid lines are linear fits to the data in order to measure its internal resistance.}
 \label{fig:plot1}
\end{figure}

\begin{figure}[h]
 \includegraphics[width=80mm]{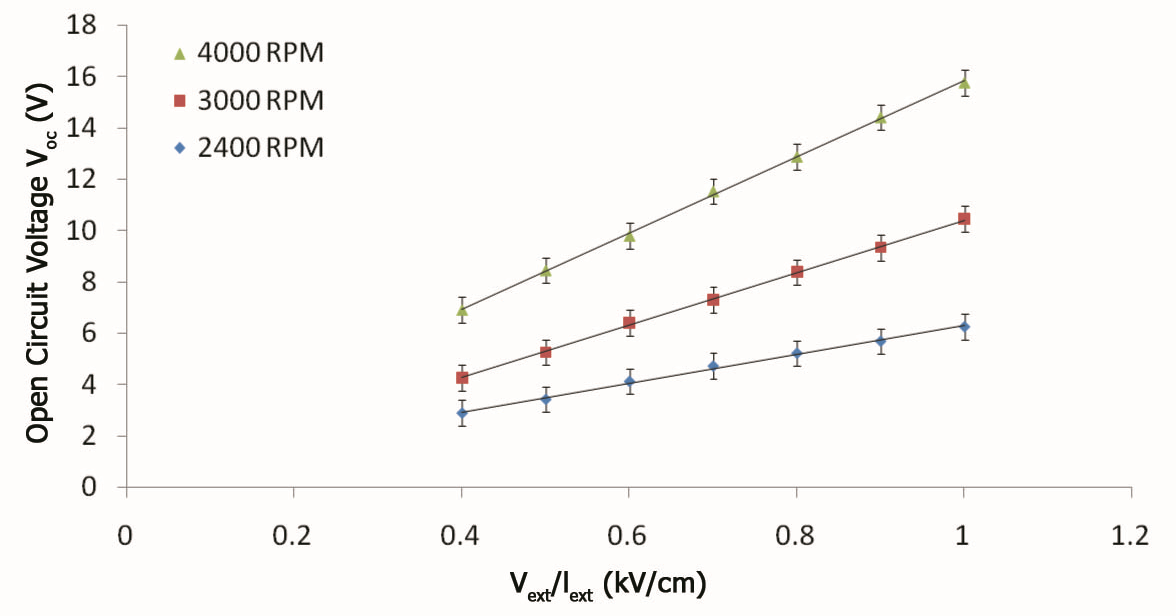}
 \caption{Generated voltage as a function of the external voltage in different angular velocities.
 	Solid lines are linear fits to the data.}
 \label{fig:plot2}
\end{figure}

\begin{figure}
\centering
\includegraphics[width=70mm]{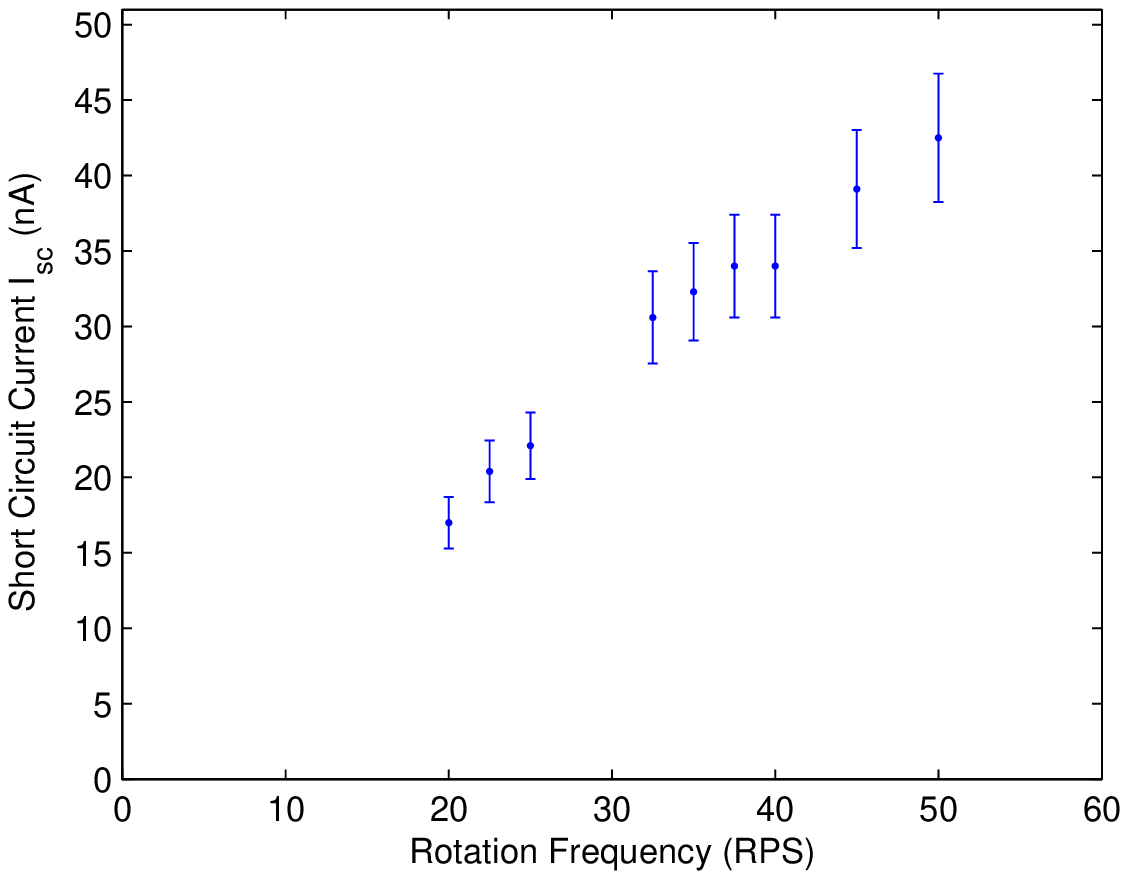}
\includegraphics[width=70mm]{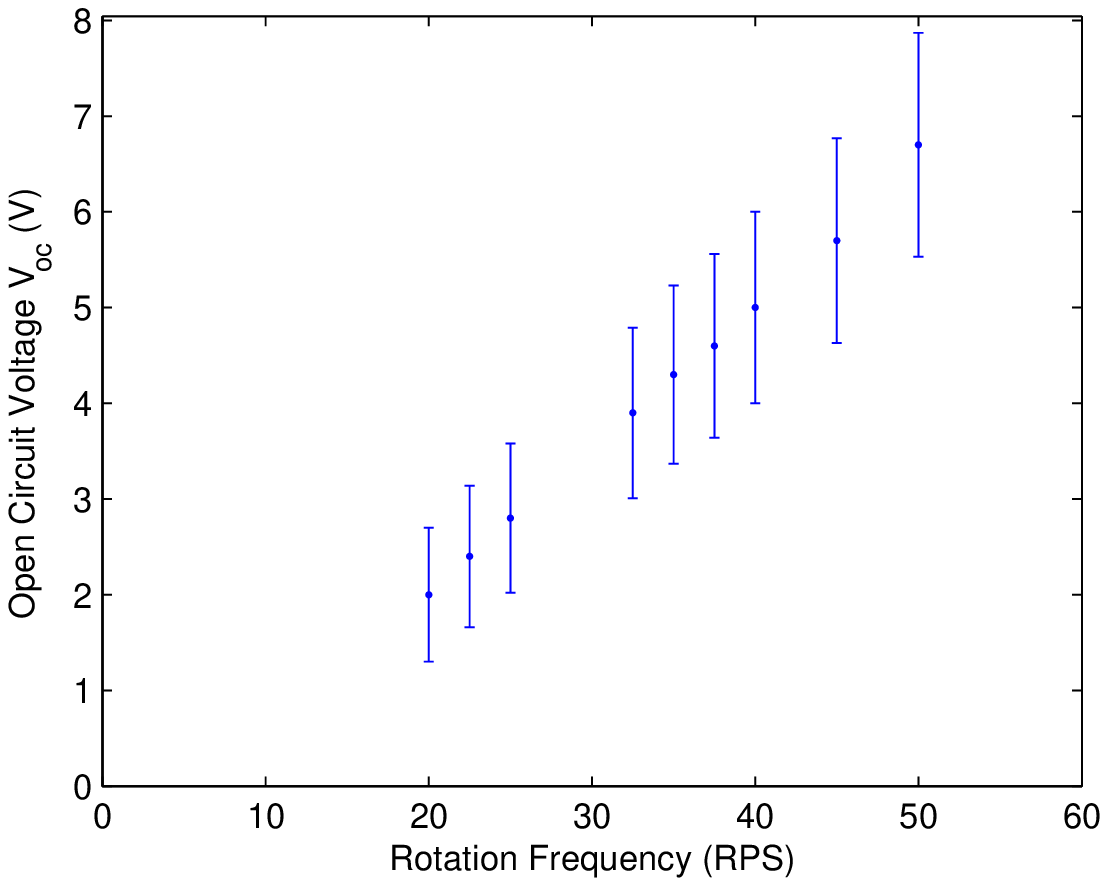}
\caption{Dependency of generated power to the rotation frequency. Top: Short circuit current. Bottom: Open circuit voltage. $V_{ext}= 6 kV$.}
\label{fig:charac}
\end{figure}

\begin{figure}[h]
 \includegraphics[width=\linewidth]{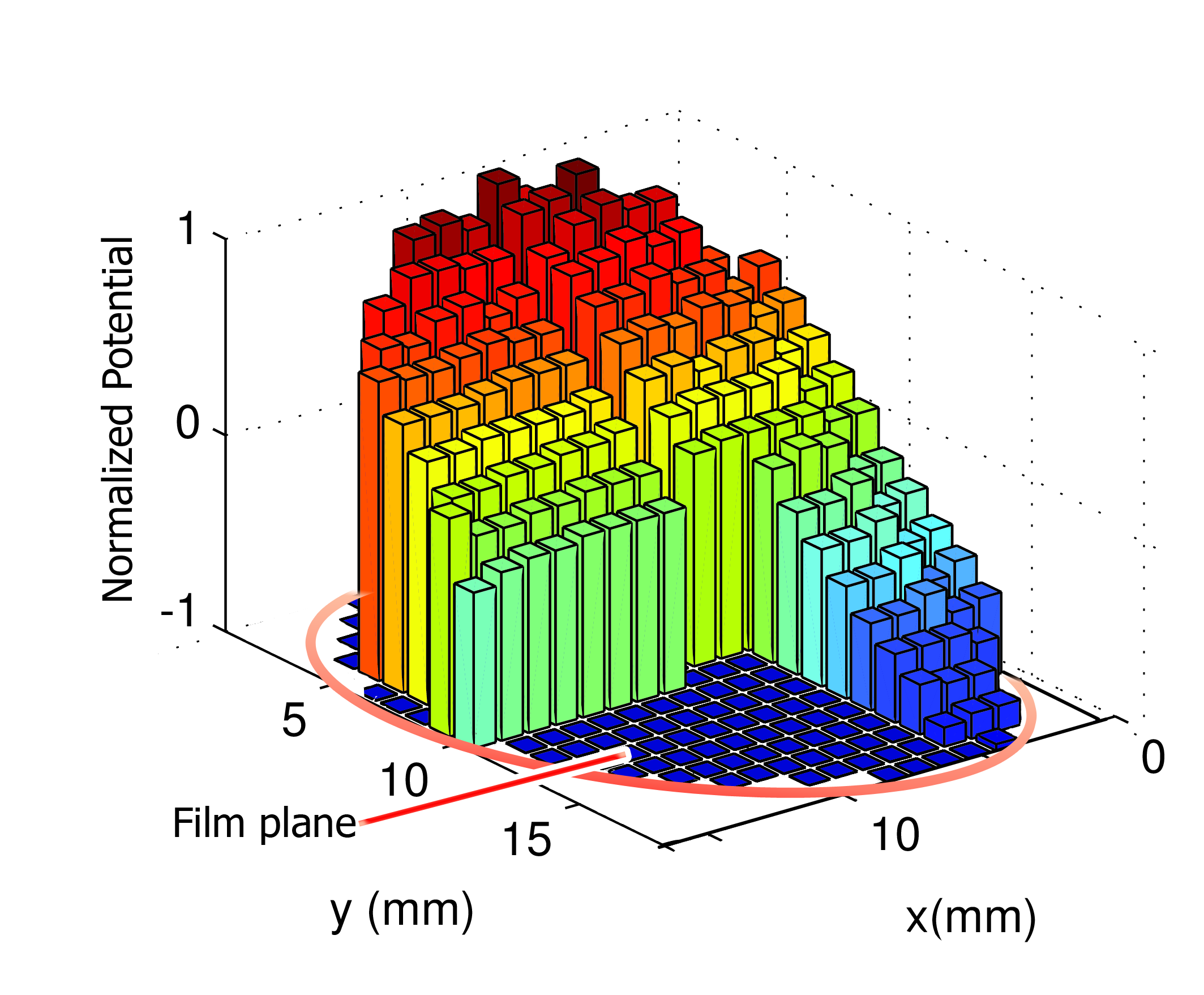}
 \caption{Measured relative potential at different points on the film rotating with an angular velocity of $\approx 2000 RPM$. The film is sketched in the x-y plane.}
 \label{fig:pot}
\end{figure}

Usually, the relation between output current vs. voltage (I-V characteristics) of an electric generator is a main standard parameter, 
by which Norton or Tevenan equivalents of the generator can be achieved (Fig. \ref{fig:nort}). The output parameter of the generator is $R_{Nor} = R_{Tev} = V_{oc} / I_{sc}$  where $R_{Nor}$ and $R_{Tev}$ are the Norton and Tevenan resistances and $V_{oc}$ and $I_{sc}$ are open circuit voltage and short circuit current respectively.
From this parameter one can find the output impedance of a generator in different conditions and also the output power of the system could be investigated under different excitations and loads.
To find the I-V characteristics of the device, the electrodes were attached to a variable load resistor ranging from $100M \Omega$ to $1100M \Omega$ with steps of $100 M \Omega$ and the voltage differences were measured using a voltmeter which had a very high input impedance of $10^{14} \Omega$ (using Keithley 602). To measure the open circuit voltage, the variable load was removed, leaving only the internal resistance of the voltmeter in the circuit. The current is obtained from measured voltage divided by the load resistance.
By testing the voltage across the system in different load resistances the I-V characteristics of the generator is obtained for different angular velocities as shown in Fig. \ref{fig:plot1}. The slope of the I-V characteristic fitted line is the inverse of the output impedance of the generator or $R_{Nor}$. This output impedance is measured to be $\approx 300M \Omega$ in 4000 $RPM$, and drops down to $\approx 150 M \Omega$ in 2400 $RPM$ which is corresponding to the thickness of the film and implies that at higher velocities, the film becomes thinner leading to a higher ohmic resistance.
The output voltage of this device for different angular velocities is plotted as a function of external voltage in Fig. \ref{fig:plot2}. A linear relation between $V_{ext}$ and the produced voltage is obtained which is verified theoretically in section 3.

The basic characteristics of the system are the open circuit voltage and the short circuit current, which are obtained from extrapolation of the I-V characteristics. The results of experiment in Fig. \ref{fig:plot1} show that there is a linear relation between the produced voltage and current. So the two basic characteristics of the system were measured experimentally as a function of the angular frequency.
Results are shown in Fig. \ref{fig:charac}.
The created voltage is in a direct relation with the angular velocity of the film i.e. reversing its direction of rotation reverses the voltage and increasing its velocity increases the output voltage (Fig. \ref{fig:charac}). 

In order to investigate the distribution of potential on the film's surface, we performed an experiment in which the potential of several points on the film were measured by an x-y micrometer adjustable electrode. Results are shown in Fig. \ref{fig:pot}, it shows that the potential is changing almost linearly as a function of $y$. The potential gradient at velocities under $4000 RPM$ is approximately zero in the $x-$direction.

\section{Discussion}

\begin{figure}
\includegraphics[width=\linewidth]{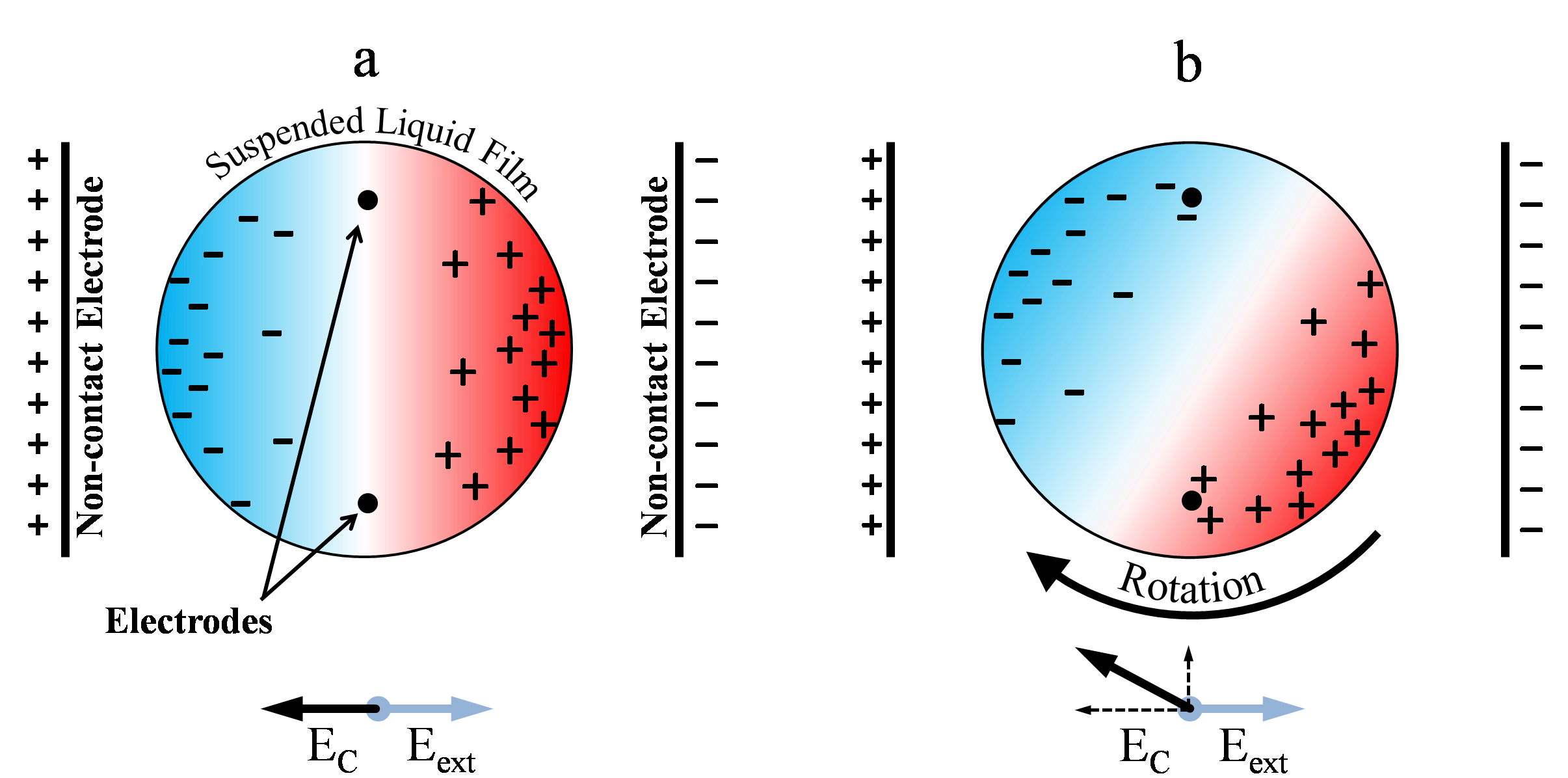}
 \caption{Schematic illustration of the liquid film electric generator mechanism. (a) The base state: Electric charge distribution on the film in the static state in which the fluid is quiescent. By this charge distribution the total electric field ($\vec{E}_C + \vec E_{ext}$) will be zero all over the film. (b) The charge distribution in a rotating frame. A gradient electric potential normal to the external field will be formed in this case.}
 \label{fig:illustration}
\end{figure}



The physical mechanism behind the liquid film electric generator is similar to electroconvection in thin films.
Recalling from the literature on electroconvection \cite{tsai2007direct}, the governing equations addressing electrohydrodynamics in a suspended thin liquid film are as below. In this context, magnetic forces and dielectric effects are negligible.
\begin{eqnarray}
\nabla \cdot \vec{u} = 0 \\
\rho \Big( \frac{\partial \vec{u}}{\partial t} + (\vec{u} \cdot \nabla)\vec{u} \Big) = 
	-\nabla P + \mu \nabla^2 \vec{u} + q \vec{E} \\
\frac{\partial q}{\partial t} = -\nabla \cdot(\sigma \vec{E} + q \vec{u}) \label{eq:4} \\
q = -2 \epsilon_0 \partial_z \psi_3 \big|_{z=0^+} \label{eq:5} \\
\nabla_3^2 \psi_3 = \nabla_2^2 \psi_3 + \frac{\partial^2 \psi_3}{\partial z^2} = 0 \label{eq:6} \\
\psi_2 = \psi_3(z=0)
\end{eqnarray}

The first two equations are the continuity and momentum of Navier-Stokes equations, in which $\vec{u}$ is the fluid velocity, $P$ is pressure, $\rho$ is surface mass density, $\mu$ is surface viscosity, $q$ is surface charge density and $\vec{E}$ is the electric field in the fluid which is related to the potential on the film by $\vec{E} = -\nabla_2 \psi_2$.
The third equation is the conservation of charge, which is transferred by an advection term $q\vec{u}$ and an ohmic conduction term $\sigma \vec{E}$ where $\sigma$ is the conductivity multiplied by the film thickness. $\epsilon_0$ is the vacuum permittivity, and subscripts denote two and three dimensional potentials and gradients.

If the film is stationary, the external voltage on the non-contact electrodes creates an electric field of $\vec{E}_{ext}$ in the $x-$direction. Since the film is conductive, charges will be induced on its surfaces to cancel out this field inside the film. As a result, in a steady state the total electric field in the film, which is a superposition of the external field and the charge field, $\vec{E}_C$, will be zero. We call this the base state, and the corresponding charge configuration is $q_0(x, y)$. When the film is in motion, the charge configuration changes because of the advection of charges, and the total charge on the film is no longer equal to $q_0$, but is $q(x, y) = q_0 + q_{loc}$, where $q_{loc}$ is the changes of charge in each point. As the charge configuration changes, the electric field is no longer zero, and a potential difference occurs on the film (Fig. \ref{fig:illustration}).

Thus it is the advection term in equation 3, $\vec{\nabla} \cdot(q \vec{u})$ that causes the charges to move with the fluid and produces the convective current. In our case, current of the liquid film electric generator can be treated as a forced convective current, i.e., the external voltage creates a charge gradient in the $x-$direction, then the forced film motion advects the charges to the two sides of the film in a way that make charge gradient in $y-$direction. 
The liquid film electric generator can be assumed as the utilization of the electroconvection phenomenon to create electric potential and current.

According to this explanation, the electricity generation is a competition between the advective term which induces a current and the conductive term which consumes the current. Thus if we increase the conductive current by increasing the conductivity of the film, the generated voltage and current intensity will drop. This fact was observed experimentally by adding salt to the solution to increase its conductivity. We clearly observed that as expected theoretically, adding a small amount of salt decreases the measured voltage.
Recalling from our previous study on the liquid film motor, we had the same experience that in constant external electric field and constant internal current intensity, adding salt to the liquid film slows down its rotation speed.


Equations \ref{eq:4}, \ref{eq:5} and \ref{eq:6} link the surface charge distribution on the film with the potential in each point. 
The external voltage appears as a boundary condition on the non-contact electrodes for the Laplace equation \ref{eq:6}.
The base
state for the charge distribution $q_0$ can be calculated assuming zero velocity and steady state.
Based on these equations, the actual relation between potential and surface charge density is non-local, i.e., the potential on each point is affected by the charge distribution all over the film. However as an approximation, a local relation can be obtained between the local charge density and potential, as done before in the case of electroconvection \cite{tsai2005charge}. In this local approximation, the potential caused by the changes of charge distribution in a specific point is directly proportional to the changes of charge density from the base state therein: $\psi_{loc} = \alpha q_{loc} = R q_{loc} / \epsilon_0$ where $R$ is the radius of the film. In the base state the potential is equal all over the film. We assume this potential to be zero. So the total potential can be calculated as a function of the charge distribution:
\begin{eqnarray}
\psi_2 = \psi_0 + \psi_{loc} = \psi_{loc} = \alpha q_{loc} = \alpha (q - q_0).
\end{eqnarray} 

To find the base state, the Laplace equation \ref{eq:5} must be solved by assuming the potential on the film to be zero, and by applying appropriate boundary conditions on the external electrodes and the contact electrodes. Using equation \ref{eq:4} the base state charge distribution will be found. To come to a simple understanding on the main mechanism, here we will assume that the base state charge distribution is linearly a function of $x$, i.e. $q_0 = -E_{ext} \cdot x / \alpha$ where $E_{ext} = V_{ext} / l_{ext}$. This is while in reality the base state has a more complex solution. As a result of this simplification:
\begin{eqnarray}
\psi_2 = x E_{ext}+ \alpha q.
\label{eq:local}
\end{eqnarray}

\begin{figure*}
\raisebox{18mm}{\includegraphics[width=40mm, angle=90]{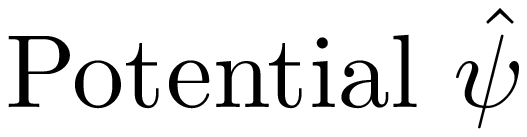}}
\includegraphics[width=39mm]{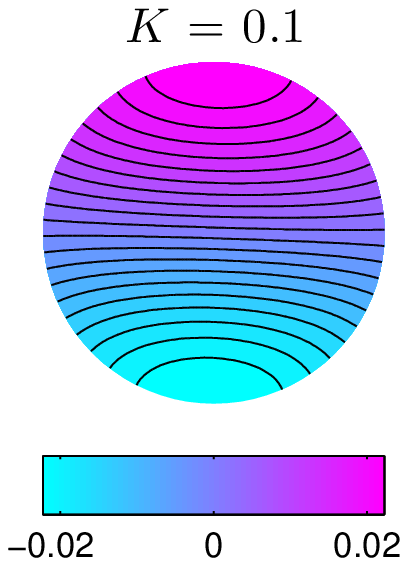}
\includegraphics[width=39mm]{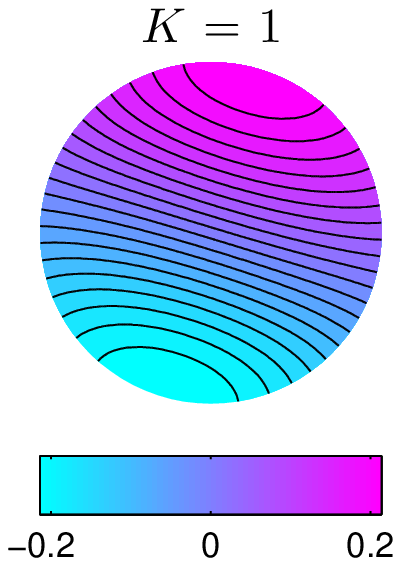}
\includegraphics[width=39mm]{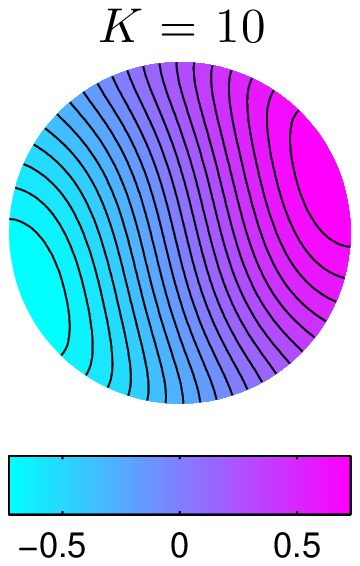}
\includegraphics[width=39mm]{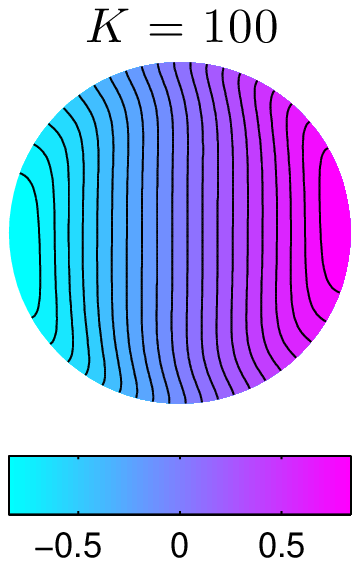} \\
\raisebox{14mm}{\includegraphics[width=40mm, angle=90]{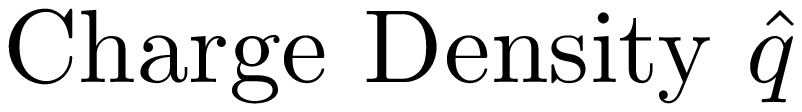}}
\includegraphics[width=39mm]{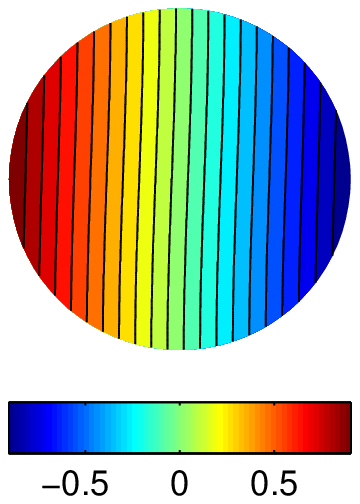}
\includegraphics[width=39mm]{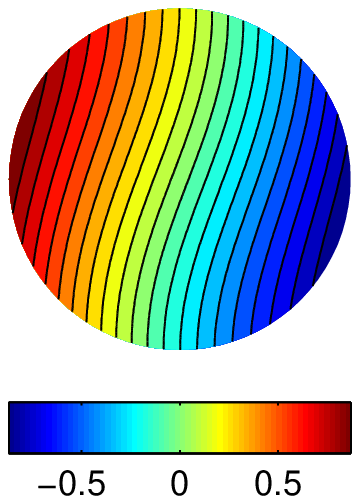}
\includegraphics[width=39mm]{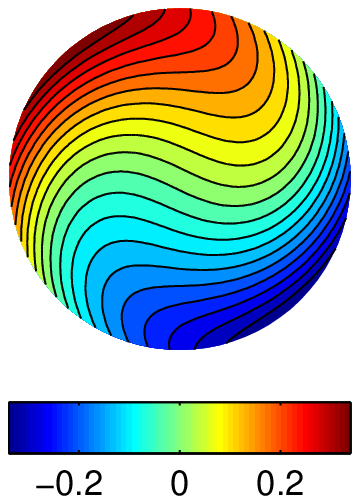}
\includegraphics[width=39mm]{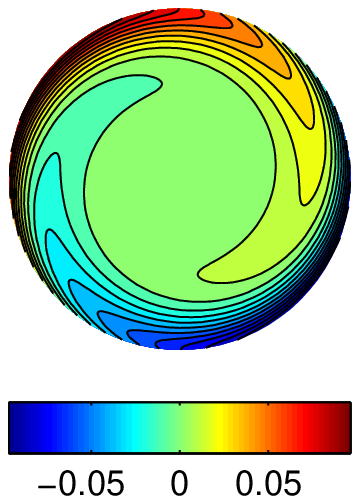}
\caption{\label{fig:sim} Contours of dimensionless potential $\hat{\psi}$ (top plots) and dimensionless charge density $\hat{q}$ (bottom plots) in different values for the velocity parameter $K = \omega R^2 / (\sigma \alpha)$: from left to right: 0.1, 1, 10 and 100. Calculated numerically with the local approximation.}
\end{figure*}

We have observed that in our range of experiments with water, the film acted almost rigid in its dynamics, i.e. the velocity was $\vec{u} = (r \omega \sin{\theta}, -r \omega \cos{\theta})$ where $\omega$ is the angular velocity of the film. This rigid behaviour is due to the high rotation velocity compared to the velocity that could be induced with the electric forces. Applying this simplification and the approximation used for equation \ref{eq:local} to the governing equations, a single linear equation will rule the phenomenon. In terms of non-dimensional parameters and polar coordinates:
\begin{eqnarray}
\frac{1}{\hat{r}} \frac{\partial \hat{\psi}}{\partial \hat{r}}+\frac{\partial^2 \hat{\psi}}{\partial \hat{r}^2} +
	K \frac{\partial \hat{\psi}}{\partial \theta} + \frac{1}{\hat{r}^2} \frac{\partial^2 \hat{\psi}}{\partial \hat{r}^2}
	= -K \hat{r} \sin{\theta} \label{eq:main}
\\
\hat{\psi}\big|_{\hat{r} = 0} = 0 \ \ \ \ \ \ \ \ \ \ \frac{\partial \hat{\psi}}{\partial \hat{r}}\big|_{\hat{r} = 1} = 0
\end{eqnarray}

Where $K = \omega R^2 / (\sigma \alpha)$ is the dimensionless velocity parameter, and the hat signs denote non-dimensional parameters: $\hat{r} = r / R$, $\hat{\psi} = \psi_2 / (E_{ext} R)$. The dimensionless charge density is defined and calculated as $\hat{q} = q \alpha / (E_{ext} R) = \hat\psi - r \cos \theta$. 
The dimensionless velocity parameter $K$ can be thought of as dimensionless angular velocity multiplied by the dimensionless charge relaxation time $\tau_C$.

We solved equation \ref{eq:main} by a discretization of angle to 100 and radius to 80 parts, having $N = 8000$ points. Thus the equation becomes a system of $N$ linear equations with the potential on each point as an unknown. Solving the equations with MATLAB, we could get the results of this equation as a function of the dimensionless number, $K$ as shown in Fig. \ref{fig:sim}.

The results show that in small angular velocities, the potential has a gradient in the $y$-direction, and a magnitude proportional to the angular velocity $\omega$. This is in agreement with our observations in experiments. As the angular velocity increases, the distribution of charges changes from the base state because the velocity of rotation will be significant in comparison with the velocity of charge conduction. In this case, the equipotential lines will eventually rotate with the increase of $\omega$. As $K$ increases, the charges
will be spread in the film and the local charge densities will decrease, so in very high velocities, the charges cannot affect the potential, and $q$ will be negligible in equation \ref{eq:local}. This pattern is clearly visible in Fig. \ref{fig:sim}.
We tried to observe this in experiments by measuring the angle between the external field and the electrodes that the maximum voltage difference occurs in. By increasing the rotation velocity over $5000 RPM$, we observed that this angle varies from the basic 90-degrees in the direction of rotation, as the numerical results indicate. However the film becomes unstable in high velocities and quantitative experimental results are difficult to obtain.

\section{Conclusion}
The liquid film electric generator (LFEG) is in contrast to the liquid film motor, both of which work with the same principle. We investigated the electrical characteristics of such a generator. It was shown that the generated voltage and current are in direct relation with the excitations of the system, which are the angular velocity and the external voltage. Increasing the conductivity of the film decreases the generated voltage, which is consistent with the electroconvective theory for the effect.

For a theoretical analysis, the local approximation for the charge-potential relation was used as well as a simplifying linear assumption for the base state, and by assuming rigid dynamical behaviour, equations of electroconvection were simplified and solved numerically to describe the phenomenon. Results are in qualitative agreement with the experiments. 

In spite of the effort of this contribution, the liquid film electric generator deserves further investigation. More
sophisticated experimental configurations should be designed to allow for better measurements, which is vital for a quantitative study. In addition, for a precise theoretical analysis, the non-local relation between the surface charge density and potential should be taken into account, also flow dynamics should be included. It is worth to mention that after understanding the LFM and LFEG, there is still more researches to be performed in order to eventually produce a suitable micro-machine.


\section*{Acknowledgement}
This work was supported by Sharif Applied Physics Center. We wish to thank S. O. Sobhani for his efforts in experiments. We also thank Professor S. W. Morris, Professor R. Shirsavar and Professor M. R. Ejtehadi for their helpful comments during this research.



\bibliographystyle{spphys}       

\end{document}